\documentstyle[prb,aps,multicol,psfig]{revtex}
  \renewcommand{\narrowtext}{\begin{multicols}{2} \global\columnwidth20.5pc}
  \renewcommand{\widetext}{\end{multicols} \global\columnwidth42.5pc}

\begin{document}  
\newcommand{\beq}{\begin{equation}}  
\newcommand{\eeq}{\end{equation}}  
\newcommand{\be}{\begin{eqnarray}}  
\newcommand{\ee}{\end{eqnarray}}  
\newcommand{\p}{\partial}  
\newcommand{\E}{{\rm E}}  
\newcommand{\W}{{\rm W}}  
\renewcommand{\Im}{\mathop {\mathrm {Im}}} 
\renewcommand{\Re}{\mathop {\mathrm {Re}}} 
\title{ 
Acoustoelectric effect in a finite-length        
ballistic quantum channel 
 } 
\author{O. Entin-Wohlman$^{(a)}$\thanks{Permanent address: School of Physics 
and Astronomy, Raymond and Beverly Sackler Faculty of Exact Sciences, 
Tel-Aviv University, Tel-Aviv 69978, Israel}, Y. Levinson$^{(a,b)}$, 
and Yu. M. Galperin$^{(a,c)}$  
}  
\address {$^{(a)}$Centre of Advanced Studies, Drammensveien 78, 0271 
Oslo, Norway;\\ $^{(b)}$Department of Condensed Matter Physics, The Weizmann 
Institute of Science, Rehovot 76100, Israel; \\$^{(c)}$Department of 
Physics, University of Oslo, Box 1048 Blindern,   
N-0316 Oslo, Norway and  
Solid State Division, \\ and A. F. Ioffe Physico-Technical Institute, 194021  
St. Petersburg, Russia }   
\date{\today}  
\maketitle  
  
\begin {abstract}  
The dc current induced by a coherent surface acoustic wave (SAW)  
of wave vector $q$ in a ballistic channel of length $L$ is calculated. 
The current contains two contributions, even and odd in $q$. 
The even current exists  only in a asymmetric channel, when 
the electron reflection coefficients $r_{1}$ and $r_{2}$ at both  
channel ends are different. 
The direction of the even current does  not depend on the direction 
of the SAW propagation, but is reversed upon interchanging 
 $r_{1}$ and $r_{2}$. 
The direction of the odd current is correlated with the direction 
of the SAW propagation, but is insensitive to the interchange of 
 $r_{1}$ and $r_{2}$.  
It is shown that  both contributions  to the  
current are  non  zero only when the electron 
reflection coefficients at the channel ends are energy dependent.   
The current  exhibits  geometric oscillations  as  function of $qL$. 
These oscillations are the  hallmark of the coherence of the SAW 
and  are completely washed out  when the current is induced by a flux 
of non-coherent phonons. 
The results are compared with those obtained previously by different 
methods and under different assumptions. 
\end{abstract}

  
\narrowtext  
\section{Introduction}  
The acoustoelectric effect is the generation of a dc electric current  
(the so-called \emph{acoustoelectric current})  
in a non-biased device by a coherent acoustic wave or a flux of 
phonons. There has been  recently a growing interest in observing 
this effect in mesoscopic structures.  
In particular, the acoustoelectric current  due to a   
surface acoustic wave (SAW)  
was investigated experimentally in a point contact (PC)   
defined in a GaAs/AlGaAs heterostructure by a  
split gate \cite{Sh1,Sh2,Ta97}.  
In mesoscopic structures one can expect to observe  
effects related to ballistic transport, when  the length of the  
PC channel $L$ is shorter than the electron mean free path $l$.  
  
The  theoretical considerations of the acoustoelectric effect  
can be divided into  two groups.  
The first one, based on a classical approach, uses the Boltzmann   
equation for the electrons, with the  acoustic wave  
 considered either as a classical coherent force \cite{Sh1,To96},  
or as a flux of non-coherent  
 quasi-monochromatic  phonons\cite{Gu96,Gu98}.  
The  classical approach for the description of  electrons  
 is valid for not very low temperatures,  
when the temperature smearing destroys the interference  
of the electron waves. For a ballistic PC  the relevant  
interference is due to reflection from the  channel  
ends, and one can use the Boltzmann equation for the electrons  
when  $T\gg v/L$, where $v$ is the relevant electron  velocity. (For 
brevity here, as well as in the following expressions, we put $\hbar =1$).  
For lower temperatures the quantum approach has to be used  
\cite{Ma97,Lev00}.  
  
The situation considered in Refs.~\onlinecite{Sh1} and  
~\onlinecite{To96} 
 does  not correspond  
to a ballistic electron propagation, since the channel was    
assumed to be infinitely long,  
which means that $qL,kL\gg 1$ and $L\gg l$, where $q$ is the   
wave vector of the SAW and  $k$ is the relevant electron  momentum.  
Nevertheless, it has been conjectured that those results can be  
carried over to the ballistic PC by  
replacing the mean free path $l$ by the channel length $L$.  
As we  show this is not totally correct.  
A ballistic situation was considered in Refs.~\onlinecite{Gu96} 
  and  ~\onlinecite{Gu98},  
where  the  SAW was represented by  a phonon flux,  
instead of   a classical force.  
It is known,( see e. g. Ref.~\onlinecite{Ga?}),   
that for  an infinite channel both    
representations of the  acoustical wave are equivalent 
for the derivation of  the acoustoelectric current.   
But, as  follows from our results, the  
SAW representation  by  a phonon flux  is not always adequate  
for a ballistic channel of a  finite length.  
  
In what follows we consider the  classical approach for a  
ballistic channel of a finite length,  
representing the SAW by  a classical force,  
and allowing for electron reflections from the channel ends.  
As has been observed in  Refs.~\onlinecite{Sh1}  
and explained in Refs.~\onlinecite{Sh1}  
and ~\onlinecite{Gu96},  
the   
acoustoelectric current is high at the thresholds  of the  channel   
openings, i.e. at the steps of the PC quantized conductance  
(giant acoustoelectric current  oscillations ).  
In this situation the current is due to ``resonant'' electrons,  
whose  velocities $v$ are of  order of the SAW velocity $s$.  
However, these  slow electrons have short mean free paths  
and their propagation in most PC's is not ballistic.  
We will calculate the acoustoelectric current which  
corresponds to the plateaus of the quantized conductance,  
i.e. far from the channel opening threshold, where   
$v\gg s$ and a ballistic electron propagation is more realistic.  
 This current appears to be  larger when  the PC  
 is not symmetric, i.e. when the reflection coefficients  
from both channel ends are different.  
   
\section{The Boltzmann equation}  
  
\begin{figure}[h]
\centerline{
\psfig{figure=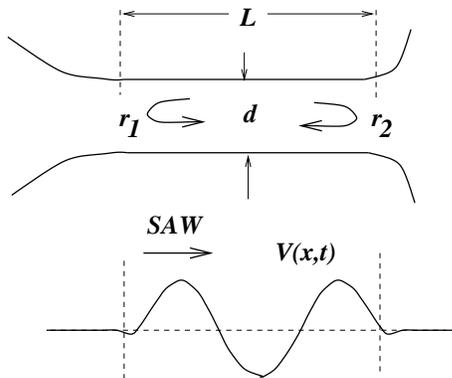,width=6cm}
}
\caption{Upper panel -- a sketch of the system under consideration. An
asymmetric 
channel of  length $L$ and width $d$ is defined in a 2DEG by a split
gate. $r_{1,2}$ are the electron reflection coefficients from the
channel's edges. Lower panel -- a snapshot of the profile of the potential
created by the SAW. Outside the channel the field is screened by the
wide leads. \label{f_1}}
\end{figure}

We consider a PC which is shaped by a split gate as a relatively   
long and uniform  channel, which opens to (non-biased) terminals, see
Fig.~\ref{f_1}. In the channel along the $x$ direction   
the electronic  states are quantized  
in the transverse direction $y$  and as a result the electron  
energy is $E_{n}+\epsilon_{k}$, where $n$ labels  the   
transverse modes  and $k$ is the electron momentum along the channel.  
$E_{n}$ is the threshold energy for the $n$th  
mode and $\epsilon_{k}=k^2/2m$ is the longitudinal electron energy.  
A mode $n$ contributes to the current if $E_{n}<E_{F}$, where $E_{F}$  
is the Fermi energy in the device terminals.  
We assume that only the lowest transverse mode $n=0$ is relevant,   
namely  
that $E_{F}$ is above the first threshold (the PC pinch-off),  
but below the next one.  
The electrons of the relevant mode are described by the  
 distribution function  $f(x,k,t)$, which satisfies  
 the Boltzmann equation. In the absence of scattering this equation is  
\be  
\label{Be}  
{\partial f\over\partial t}+v  
{\partial f\over\partial x}-  
{\partial V\over\partial x}{\partial f\over\partial k}=0,  
\ee  
where $v=\p\epsilon_{k}/\p k$ is the electron velocity and   
$V$ is  the potential created by the SAW,   
\be  
V(x,t)=V_{\omega}\exp(iqx-i\omega t)+\text{c. c.},  
\ee  
with  $\omega=sq$. We assume the SAW potential to be totally  
screened in the terminals and not screened inside  the channel.
(For a more detailed consideration of the potential and its
screening see Refs. \onlinecite{GA98} and \onlinecite{AG98}).  
The boundary conditions at the edges of the channel   
at $x=0,L$ are   
\beq   \label{bq}  
f(k)=\left\{ \begin{array}{ll}
t_{1}f_{0}(k)+r_{1}f(-k), & (x=0,\;k>0)\;,\\
t_{2}f_{0}(k)+r_{2}f(-k), & (x=L,\;k<0)\;.
\end{array} \right.
\eeq  
Here  
$t_{1,2}$ and $r_{1,2}$ are the transmission and the reflection  
coefficients for electrons approaching the ends of the channel,  
 which satisfy the normalization $t_{1,2}+r_{1,2}=1$ and  
 depend on the electron energy $\epsilon_{k}$. In Eq. (\ref{bq}),  
\be  
 f_{0}(k)=f_{T}(\epsilon_{k})=   
\left[\exp\left({\epsilon_{k}+E_0-E_F 
\over T}\right)+1\right]^{-1} 
\ee  
is the equilibrium electron Fermi  
distribution in the absence of the SAW.  
The terms with $f_{0}$ represent electrons penetrating from the  
terminals into the channel, and the terms with $r_{1,2}$  
describe electrons  back-scattered from the terminals.  
  
For a weak SAW,  Eq.~(\ref{Be}) can be solved by expanding the  
distribution function in powers of the SAW potential $V$,  
\be  
\label{fexp}  
f(x,k,t)=f_{T}(\epsilon_{k})+f_{1}(x,k,t)+f_{2}(x,k,t)+...\; .  
\ee  
The linear response $f_{1}$ contains Fourier components  
with frequencies $\pm\omega$, while $f_{2}$ contains the   
frequencies $0$ and $\pm 2\omega$. The dc part of $f_{2}$ is   
the one yielding the dc  acoustoelectric current.  
  
Expanding the Boltzmann equation (\ref{Be}) and the boundary   
conditions (\ref{bq}) we find the  equations for $f_{1,2}$,  
\be  
\label{Be1}  
\left({\p\over \p t}+v{\p\over \p x}\right)f_{1}&=&  
-{\p V\over\p x}\left(-{\p f_{T}\over\p\epsilon}\right)v,  
\ee  
\be  
\label{Be2}  
\left({\p\over \p t}+v{\p\over \p x}\right)f_{2}&=&  
{\p V\over\p x}{\p f_{1}\over\p k},  
\ee  
and the boundary conditions, 
\beq  
\label{bq12}  
f_{1,2}(k)= \left\{ \begin{array}{ll}
r_{1}f_{1,2}(-k), & (x=0,\;k>0)\;, \\
r_{2}f_{1,2}(-k), &(x=L,\;k<0)\;.
\end{array}\right.
\eeq  
Expressing  the linear response in Eq. (\ref{Be1}) as  
\be  
\label{lr}  
f_{1}(x,k,t)=e^{-i\omega t}f_{\omega}(k,x)+\text{c. c.},  
\ee  
we find the equation for the linear response amplitude,  
\be  
\label{Belr}  
\left(-i\omega+v{\p\over\p x}\right)f_{\omega}(k,x)=  
-iqV_{\omega}e^{iqx}\left(-{\p f_{T}\over\p\epsilon}\right)v.  
\ee  
Introducing Eq. (\ref{lr})  into Eq. (\ref{Be2})  
and averaging the latter  over time,  we find  
the equation for  the dc contribution of  $f_{2}$,  
\be  
\label{Bedc}  
v{\p \overline{f_{2}}\over\p x}=2 q {\p\over\p k}\Im \left[ 
V_{\omega}^{*}e^{-iqx} f_{\omega} \right].  
\ee  
Integrating the last equation over $k$ and noting  that  
the right hand side is a full derivative with respect to $k$,  
one can see  that the time averaged acoustoelectric current  
\be  
\overline{j}=e\int_{-\infty}^{+\infty}{dk\over 2\pi}v\overline{f_{2}},  
\ee  
is constant along the channel.  
  
Solving Eq. (\ref{Belr}) one finds  
\be  
\label{sf1}  
f_{\omega}(k,x)=-iqLV_{\omega}\left(-{\p f_{T}\over\p\epsilon}\right)  
F(k,x),  
\ee  
with  
\be  
F(k,x)=e^{ipx}\left[A(k)+{e^{i(q-p)x}-1\over i(q-p)L}\right],  
\ee  
where $p\equiv \omega/v$.  
The integration constant $A(k)$ is determined  by the boundary  
conditions 
(\ref{bq12}) for $f_{1}$ giving, for $k>0$,  
\be  
\label{A}  
A(k)={e^{2ipL}r_{1}r_{2}\phi_{k}-r_{1}\phi_{-k}  
\over 1-e^{2ipL}r_{1}r_{2}},\qquad A(-k)={A(k)\over r_{1}},  
\ee  
with  
\be  
\phi_{k}={e^{i(q-p)L}-1\over i(q-p)L}.  
\ee  
Introducing Eq. (\ref{sf1}) into Eq. (\ref{Bedc})  
we find  
\be  
 v{\p \overline{f_{2}}\over\p x}&=&-q^2|V_{\omega}|^2 L 
\nonumber \\ && \times
\Re \left\{ 
e^{-iqx}{\p\over\p k}\left[  
\left(-{\p f_{T}\over\p\epsilon}\right)F(k,x)\right]\right\}.  
\label{Bedcm} 
\ee  
Solving Eq. (\ref{Bedcm}) we have  
\begin{eqnarray}  
v\overline{f_{2}}(k,x)&=&-(qL)^2|V_{\omega}|^2  
\nonumber \\ && \times
\left\{{\p\over\p k}\left[\left(-{\p f_{T}\over\p\epsilon}\right)  
G(k,x)\right] 
+B(k)\right\},  
\label{f2} 
\end{eqnarray} 
where  
\beq  
G(k,x)=2\int_{0}^{x}{dx\over L}\Re \, [e^{-iqx}F(k,x)]\, .  
\eeq  
The integration constant $B(k)$ is determined  by the boundary   
conditions (\ref{bq12})  
for $f_{2}$, to which  enters $\overline{G}(k)\equiv G(k,L)$.  
Using Eq. (\ref{A}) one finds after lengthy  but straightforward  
calculations, for $k>0$,  
\be  
\label{G}  
\overline{G}(k)&=& 
{1-r_{1}r_{2}\over|1-r_{1}r_{2}e^{2ipL}|^2} 
\nonumber \\ && \times
[(1+r_{1}r_{2})|\phi_{k}|^2-2r_{1}C_k\cos pL], 
\nonumber  \\ 
\overline{G}(-k)&=&-  
{1-r_{1}r_{2}\over |1-r_{1}r_{2}e^{2ipL}|^2}  
\nonumber \\ && \times
[(1+r_{1}r_{2})|\phi_{-k}|^2-2r_{2}C_{k}\cos pL],  
\ee  
where  
\beq  
C_{k}=2(\cos pL-\cos qL)/(q^2-p^2)L^2.  
\eeq   
  
Using Eq. (\ref{f2}) for $x=0,L$ in the boundary conditions  
(\ref{bq12}) for $f_{2}$ we find for $k>0$  
\beq  
B(k)={r_{1}(r_{2}\psi_{k}+\psi_{-k})\over 1-r_{1}r_{2}},  
\qquad  B(-k)=-{B(k)\over r_{1}},  
\eeq  
where  
\be  
\psi_{k}={\p\over\p k}\left[\left(-{\p f_{T}\over\p\epsilon}\right)  
\overline{G}(k)\right].  
\ee  
  
\section{The acoustoelectric current}  
To obtain a symmetric expression for the current it is convenient  
to calculate it as the average of its values  
at $x=0$ and $x=L$, which is  
\begin{eqnarray}  
\label{cg}  
&&\overline{j}=-{e\over 4\pi}|V_{\omega}|^2(qL)^2\int_{0}^{\infty}dk  
\nonumber \\ &&  \times \left\{  
{(1-r_{1})(1-r_{2})\over 1-r_{1}r_{2}} 
{\p\over\p k}  
\left[\left(-{\p f_{T}\over\p\epsilon}\right)  
\left(\overline{G}(k)+\overline{G}(-k)\right)\right] \right.  
\nonumber \\  && \left. \quad +  
{r_{1}-r_{2}\over 1-r_{1}r_{2}}  
{\p\over\p k}  
\left[\left(-{\p f_{T}\over\p\epsilon}\right)  
\left(\overline{G}(k)-\overline{G}(-k)\right)\right] \right\}.  
\end{eqnarray}  
One can calculate the integral by parts, using that  
$\overline{G}(\pm k)$ is zero at $k=0$ and   
$(-\p f_{T}/\p\epsilon)$ is zero at $k=\infty$.  
Strictly speaking, for electrons with  $k=0$  
the inequality $L \gg l$ required for the  channel to be    
ballistic may be not fulfilled. In this case 
the above  solution of the Boltzmann equation  
is not valid. However, in any case the contribution from $k=0$  
is exponentially small due to the factor $(-\p f_{T}/\p\epsilon)$,  
for  $E_F-E_{0}\gg T$. After integration by parts, one can see  
that the acoustoelectric current vanishes for an open channel  
($r_{1}=r_{2}=0$) or when the electron reflection coefficients  
 from the channel ends  are energy independent.  
This is a specific property of a ballistic channel.  
  
One may  separate the acoustoelectric current given by Eq. (\ref{cg})  
into two contributions,  
the \emph{even} current, $\overline{j}_{e}$, 
 and the \emph{odd} current, $\overline{j}_{o}$. 
The even current  does not change its 
sign upon the replacement $q \rightarrow -q $, 
but changes  the sign upon interchanging  $r_{1}$ and $r_{2}$. 
The odd  current  reverses  its sign upon the replacement  
$q \rightarrow -q $, but is symmetric with respect to  
$r_{1}$ and $r_{2}$. 
The correlation between the change of the propagation direction 
of the SAW and the interchange of the channel ends follows 
from obvious symmetry considerations. 
Note that only the odd current exists in a symmetric PC,  
as well as in any homogeneous medium.  
  
Simpler expressions for the acoustoelectric current   
can be given far from the threshold.  
The scattering time for  electrons in a high quality 2DEG   
can be taken to be $\tau=30$ ps, which corresponds, for  
a Fermi velocity  $v_{F}=3\times 10^{7}$ cm/s to a mean free  
path $l=10\, \mu$m. This means that channels with $L\alt 3 \, \mu$m  
are ballistic far from the threshold. The situation near the threshold 
where $v \simeq s$ is less clear. Assuming that $\tau$ is 
velocity-independent and using  $s=3\times 10^{5}$ cm/s,  
we estimate the mean free path as $l=0.1 \, \mu$m,  
which means that for resonant electrons most channel-shaped PC's  
with $L \simeq 1\, \mu$m are non-ballistic. However, the specific 
velocity dependence of the relaxation rate in quasi-1D channels is not 
known. 
 
For estimates one can  use $L$ between $1\, \mu$m and   
$10\, \mu$m,   $\omega/2\pi$ between 100 MHz and 1 GHz  
and temperature $T=1$ K.  
The change of $pL$ in $\overline{G}(\pm k)$  
within the thermal smearing defined by   
$(-\p f_{T}/\p\epsilon)$ is $(T/\epsilon_{F})(s/v_{F})qL$.  
This is small even for the highest frequencies and the longest 
channels available at present time.  
Hence one can replace the integration over $k$ by taking the  
integrand at $k=k_{F}$.  
Then  $pL\simeq \omega L/v_{F}$  and $p/q\simeq s/v_{F}$   
are also small.  
  
Based on these estimates we can  simplify the   
expressions for the even and  odd  currents using  
 $pL\ll 1$ and $p\ll q$. The results are  
\be  
 \overline{j}_{e}&=&(e|V_{\omega}|^2/\pi)  
\;\beta\;[1-\cos qL], \label{rect} \\  
 \overline{j}_{o}&=&{\rm sign}\, q\;(e|V_{\omega}|^2/\pi)\;\alpha\; 
(s/ v_{F}) 
\nonumber \\ && \times 
[2(1-\cos qL)-qL\sin qL], \label{drag}  
\ee  
where   
\be  
\beta&=&\left[{(1-r_{1})(1-r_{2})\over 1-r_{1}r_{2}}\right]^2  
{\p\over\p\epsilon}\;{r_{1}-r_{2}\over(1-r_{1})(1-r_{2})},  
\nonumber \\  
\alpha&=&{1+r_{1}r_{2}\over 1-r_{1}r_{2}}\;  
{\p\over\p\epsilon}\;{(1-r_{1})(1-r_{2})\over1-r_{1}r_{2}},  
\ee  
 are calculated at the Fermi energy.  
The odd current is smaller by a factor $s/v_{F}$  compared to  
the even one.

For short channels with $qL\ll 1$ one  finds $\overline{j}_{e}\sim 
(qL)^2$ and    
$\overline{j}_{o}\sim (qL)^4$, while for $qL\agt 1$  
 both contributions to the acoustoelectric  
current show geometric oscillations  
as  function of the product $qL$  ($qL$-oscillations), 
similar to the result of the quantum consideration \cite{Lev00}.  
 These  oscillations are  
due to the coherence  of the SAW and result from the  
compensation   
 of the forces  acting on the electrons in different  
half-waves of the SAW.   
No $qL$-oscillations are predicted when  the acoustic  
wave is described as a non-coherent flux of phonons \cite{Gu98},  
in which case  $\overline{j}$ is simply proportional to $L$. 
 
Finally, let us specify the expression for a symmetric channel  with 
$r_{1}=r_{2}=r$  which 
is convenient for comparison with  the results obtained by previous 
approaches. 
After integration by parts, Eq. (\ref{cg})  reduces to  
\be 
\label{sch}  
\overline{j}&=&{e\over 4\pi}|V_{\omega}|^2 (qL)^2  
\int_{0}^{\infty}dk\left(-{\p f_{T}\over 
\p\epsilon}\right)(|\phi_{k}|^2-|\phi_{-k}|^2)\nonumber \\ &&\times 
{1-r^4\over |1-r^2e^{2ipL}|^2}\;  
{\p\over\p k}\;{1-r\over 1+r}.  
\ee  
Disregarding   the fact that for low energy electrons realistic  
channels may be non ballistic, we can use  Eq. (\ref{sch}) 
to compare the values of the acoustoelectric current  
at the conductance plateaus, i.e. between the thresholds  
for  $\epsilon\simeq \Delta$,  (where $\Delta$ 
is a typical distance between the thresholds), 
and at the conductance steps, i.e. near the  thresholds 
for  $\epsilon\simeq ms^2$. 
To estimate the reflection coefficient and its derivative  we 
use the results in the Appendix, assuming that the channel 
opening are not adiabatic. 
At the plateau we find the following order of magnitude 
estimates: $r\simeq 1$ and  
$\p r/\p \epsilon \simeq  \Delta^{-1}$, 
while at the step we use Eq.~(\ref{rapp}). 
Replacing the Fermi function derivative by a delta function 
and assuming $qL\simeq 1$,  we obtain at the plateau 
\beq 
\overline{j} \simeq (e|V_\omega|^2/\Delta)\;(s/v_F)\, , 
\eeq 
(with $mv_{F}^2\simeq \Delta$) and at the step 
\beq 
\overline{j} \simeq (e|V_\omega|^2/\Delta)\, . 
\eeq 
The current at the plateau is smaller by the  
factor $(ms^2/\Delta)^{1/2}$   
compared to that at the step. 
 
\section{Comparison with previous approaches}  
  
The aim of this section is a more detailed comparison   
of  our results with the results  
obtained in Refs.~\onlinecite{To96} and \onlinecite{Gu98}. For 
simplicity we confine to  a symmetric channel.  
The result for a (non ballistic) infinite channel obtained in  
Ref.\onlinecite{To96} can be expressed  as follows  
\beq  
\label{Ga}  
\overline{j}={2e |V_{\omega}|^2 \over \pi s} 
\int_{0}^{\infty}\! 
\frac{dv 
\,(v/s)^2}{\left(1-v^2/s^2\right)^2+(\omega\tau)^{-2}}\left(-{\p 
f_{T}\over \p\epsilon}\right) \, . 
\eeq 
The integrand is an overlap of two peaks, centered at $v=v_{F}$  
and $v=s$. Hence the current is  maximal  when $v_{F}=s$,  
i.~e. near the channel opening threshold.  
When  the scattering is weak, $\omega\tau\rightarrow\infty $,   
only resonant electrons with $v=s$ contribute to the current,  
which diverges as $\omega\tau$.  
  
Since Eq. (\ref{Ga}) makes  sense only when $qL\gg 1$,  
 consider the limit $L\rightarrow\infty$  
in Eq. (\ref{sch}). Using the  representation of  
a delta-function  in this limit we have  
\beq  
|\phi_{\pm k}|^2={2[1-\cos(q\pm p)L]\over (q\pm p)^2L^2}\rightarrow  
\frac{(2/L)^2}{ (q\pm p)^2+(2/L)^2 }.   
\eeq  
For $qL\gg 1$ the term with $|\phi_{-k}|^2$ does not contribute to  
the integral in Eq. (\ref{sch}) and we obtain from this equation  
\be  
&&\overline{j}={2e|V_{\omega}|^2 \over \pi s} 
\int_{0}^{\infty}\!\frac{dv\, (v/s)^2\, R\{r\}}{(1-v^2/ 
s^2)^2+(4v/\omega L)^2 }  
\left(-{\p f_{T}\over \p\epsilon}\right);\nonumber \\ 
&&R\{r\}=  
{2(1-r^4)\over |1-r^2\exp(2i\omega L/v)|^2}  
\;ms\;{\p\over\p k}\;{1-r\over 1+r}.  
\label{schm} 
\ee 
Comparing Eq. (\ref{schm}) and Eq. (\ref{Ga}) one can see that  
in the resonance factor, indeed, as  conjectured, the scattering   
time $\tau$ is replaced (up to a numerical factor)  
by the life time  $L/v$ of the electron in the open channel.  
However the result given by Eq. (\ref{Ga}) misses the   
crucial factor $R\{r\}$, which combines two features:  
(i) When there is no reflection, or for an energy-independent  
reflection, the acoustoelectric current wanishes.  
(ii) $qL$-oscillations resulting from the factor   
$|1-r^2\exp(2i\omega L/v)|^{-2}$.  
Within the resonance factor width the exponent oscillates  
once, and hence some residual $qL$-oscillations are present  
in the current given by Eq. (\ref{schm}).

Next we  compare our results with those  
 obtained in Ref.~\onlinecite{Gu98},  
where the acoustoelectric  
current is due  to a monochromatic flux of phonons   
with frequency $\omega=sq$ and the electron-phonon interaction  
is presented as an electron-phonon  
collision   term in the Boltzmann equation for the electrons.  
The comparison can be accomplished  
 only up to some $q$-dependent  factors:  Ref.~\onlinecite{Gu98}  
considers  bulk phonons, for which, as shown   
in  Ref.~\onlinecite{Kn97}, the  electron-phonon interaction matrix  
elements  
have a different $q$-dependence, compared to surface phonons.

The properties of the acoustoelectric current obtained  
in  Ref.~\onlinecite{Gu98} are dominated by the energy-momentum  
conservation in the  electron-phonon collision term.  
As a result only electrons with $k_{\pm}=ms\pm q/2$  
contribute to the current: an electron with $k_{-}$ can absorb  
a phonon from the flux, being excited into $k_{+}$,  
while an  electron with $k_{+}$ can omit  a phonon  
to the flux, being de-excited into $k_{-}$.  
It is now obvious  that the current is large  
only near the threshold of channel opening, when $v_{F}$  
is of  order  $s$  (for  $\hbar\omega\alt ms^2$, which is   
usually the case). Most importantly, this collision  
 picture is valid only when  the phonon energy is larger  
than the energy uncertainty of the electron state due to the  
finite escape  time from the channel, i.e. when   
$\omega\gg v/L$.  
For resonant electrons with $v\simeq s$  
it  means $\omega\gg s/L$ or, equivalently, $qL\gg 1$.  
This latter condition, together with $kL\gg 1$, allows  
the description of the   electron-phonon interaction as a  
{\it local} scattering term in the Boltzmann equation, as  used in  
Ref.~\onlinecite{Gu98}.  
  
Comparing our results with those of  Ref.~\onlinecite{Gu98}  
one has to bear  in mind that the description of the acoustic   
wave as a classical force in the Boltzmann equation  
for electrons  is valid only when  $q\ll k$   
and $\hbar\omega\ll \epsilon_{k}$. The latter condition  
allows to  neglect the discreteness of quantum transitions  
for phonon emission or absorption.   
For resonant electrons with $v\simeq s$ both these conditions  
 hold as long as  $\hbar\omega\ll ms^2$ or, equivalently,  
$q\ll ms$.   
  
The  flux of phonons can be considered as   
 non coherent when  the channel is longer  
than the phonon coherence length $\Lambda=s/\delta\omega$, where  
$\delta\omega$  ( smaller than the central frequency $\omega$)  
 is the width of the phonon distribution or the  SAW spectral width.  
The  condition $\Lambda\ll L$  
can be  fulfilled only when  $qL\gg 1$ and   
we may begin the comparison using  
our result in the form of Eq. (\ref{schm}).  
To imitate the non-coherence of the phonons we average  
this equation over $\omega$ with a Lorentzian form factor  
having $\delta\omega$ as half width at half height. 
This corresponds to a phonon field which can be obtained, 
for example, by selecting from  a thermal phonon distribution 
those phonons, which have frequencies within $\delta\omega$ 
and wave vectors along the channel direction.  
 In the time  representation it corresponds a quasi-monochromatic wave 
 with  fluctuating amplitude and phase. 
  The only factor to be averaged is $|1-r^2\exp(2i\omega L/v)|^{-2}$,  
since the change of the width in  the resonance factor is small.  
The result of the averaging is   
\beq  
{1\over 1-r^4}\;  
{1+\eta\cos\varphi\over 1+2\eta\cos\varphi +\eta ^2},  
\eeq  
where  $\varphi=2\omega L/v$ (with the central frequency  
$\omega$) and $\eta=r^2\exp(-2\delta\omega L/v)$  
(with $v\simeq s$).  
For strong  decoherence, $\Lambda\ll L$, one finds $\eta$   
to be exponentially small, and the $qL$-oscillations are  
totally suppressed.  
  
After this averaging, having in mind that for the comparison we are  
interested in the case $qL\gg 1$,  
the resonance factor in Eq. (\ref{schm}) can be replaced by a   
delta-function, giving  
\be  
\label{schmm}  
\overline{j}=e|V_{\omega}|^2 qL  
\left(-{\p f_{T}\over \p\epsilon}\right)_{v=s}  
\left[{1\over 2}ms{\p\over\p k}{1-r\over 1+r}  \right]_{v=s}.  
\ee  
This result has to  be compared with Eq.~(27) of Ref.~\onlinecite{Gu98},   
which corresponds to phonon momenta $q$  
below the  \u{C}erenkov threshold $q=2ms$. 
The current given by Eq.~(15) of Ref.~\onlinecite{Gu98}, 
 for phonon momenta $q$ above the threshold, 
 cannot be compared  
with Eq.~(\ref{schmm}), since our approach, as mentioned above,   
overlaps with the approach used in  Ref.~\onlinecite{Gu98} only when  
$q\ll ms$.  
  
In Eq.~(27) of Ref.~\onlinecite{Gu98} one finds the difference of the Fermi  
distributions at $k=ms\pm q/2$, which for  $q\ll ms$ reproduces  
the derivative  of the equilibrium distribution in Eq.~(\ref{schmm}).  
The acoustoelectric current is proportional to $L$ in both  
cases.  
However, there is a difference regarding the role of  
reflection from the channel ends.  
  
According to  Ref.~\onlinecite{Gu98}, above the \u{C}erenkov threshold  
the acoustoelectric current is  nonzero even for an open channel, $r=0$.  
As already mentioned, this does  not contradict Eq.~(\ref{schmm}). 
Below the \u{C}erenkov threshold, the  current  
is nonzero only in the case of an energy dependent  reflection,  
which agrees  with  Eq.~(\ref{schmm}).  
Nevertheless, the details of the derivative function in this equation  
and in Eq. (27) of Ref.~\onlinecite{Gu98} are different.  
 
The physical origin  of the discrepancy is as follows.  
In the Boltzmann equation used  in Ref.~\onlinecite{Gu98} the   
electron-phonon collision  term is local in space and is separated from  
the boundary conditions, which are  also local.  
The locality of the collision  term  is  determined 
 by the largest of the two wavelengths, $2\pi/k$ and $2\pi/q$,  
while the locality of the boundary condition is determined  by  
$2\pi/k$. They can be separated only when they are of the same order,  
i.e. when $q\simeq k$. However, when $q\ll k$ the electron-phonon  
scattering is nonlocal near the boundary.  
Formally,  the right hand side of Eq.~(4) in  
Ref.~\onlinecite{Gu98} for the nonequilibrium part of the electron  
distribution function  contains  
no information about the boundary conditions, while the  
right hand side of Eq.~(\ref{Bedcm}) depends on the boundary conditions,  
because  the linear response $f_{\omega}$ is sensitive to them.

As a result we have to conclude, that the acoustoelectric currents  
produced in a ballistic channel by a coherent SAW or  a flux  
of non-coherent quasi-monochromatic phonons are different.  
Even after destroying the coherence of the SAW the results  
are different.  
{\it A priori} there are no reasons why both representations  
of the SAW have to give the same results.   
  
\acknowledgements 
We acknowledge useful discussions with  V. I. Kozub. 
This research is partially  supported by grants from the Israeli
Science Foundation and from the Israeli Ministry of Sciences and the
French Ministry of Research and Technology.

\appendix  
  
\section{Reflection at the end of a channel} 
 
\begin{figure}[t]
\centerline{
\psfig{figure=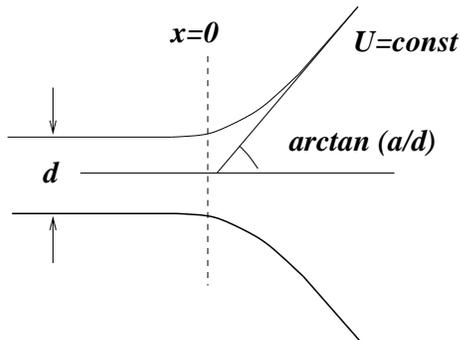,width=6cm}
}
\caption{Opening of a semi-infinite channel of width $d$ into a horn
with the opening angle defined by the ratio $a/d$. Shown are the
equipotential lines.  \label{f_2}}
\end{figure}

To have an explicit  expression for the reflection coefficient $r$ 
entering the boundary condition for the electron distribution   
function,  we 
calculate the  reflection coefficient from the end of a channel. 
For that we use the  following model for the confining potential, see
Fig.~\ref{f_2}.  
\beq  
U(x,y)={1\over 2md^{2}}\left\{ \begin{array}{ll} 
[-(x/a)^2+(y/d)^2],& (x>0)\\   
(y/d)^2, &(x<0) \end{array}\right. \, . \label{sp}  
\eeq 
This potential describes a semi-infinite waveguide at $x<0$ 
of  width $d$ connected to a horn with an opening angle $d/a$ at $x>0$. 
The variables $x$ and $y$ can be separated and the solution of the  
wave equation is of the form $\Phi_{n}(y)\psi(x)$, see Refs. 
\onlinecite{Lev00}  and \onlinecite{Le92}. 
Here $\Phi_{n}$ is a normalized harmonic oscillator wave function 
with energy $E_{n}=\Delta(n+1/2)$, where  
$\Delta=1/md^2$, and $n=0,1,2,...$ labels the modes in the waveguide 
and in the horn. There is no channel mixing upon wave transmission 
and reflection from the waveguide to the horn and vice versa. 
 
To calculate the reflection amplitude $R$ for the $n$th  mode 
for an electron energy $E$ we choose the 
wave function in the waveguide at $x<0$ as  
\be 
\psi(x)=A(e^{ikx}+Re^{-ikx}), k=[2m(E-E_{n})]^{1/2}, 
\ee 
and in the horn at $x>0$ as 
\beq 
\psi(x)=B \E (-\varepsilon,\xi), \xi=(2/dL)^{1/2}x, 
 \varepsilon=(E-E_{n})/\delta, 
\eeq 
where $\E$ is the complex Weber (parabolic cylinder) function 
 as defined in Ref.~\onlinecite{Ab64} and $\delta=1/mad$. 
(We use the notation $\E$ instead of $E$ to avoid confusion 
with the energy). $\varepsilon =0$ corresponds to the threshold of 
the $n$th mode opening. 
 As defined,  
$\psi(x)$ at $x>0$ has only a wave propagating to $x=+\infty$. 
It is convenient to use the representation, see Ref.~ \onlinecite{Ab64}, 
$$\E (-\varepsilon,\xi) 
=\sigma(\varepsilon)^{-1/2}\W (-\varepsilon,\xi)+ 
i\sigma(\varepsilon)^{1/2}\W (-\varepsilon,-\xi),$$  
where $\W$ are the real Weber functions and  
$$\sigma(\varepsilon)=(1+e^{-2\pi\varepsilon})^{1/2}-e^{-\pi\varepsilon}.$$

Matching the logarithmic derivative at $x=0$ one finds 
\be 
\label{i} 
{1-R\over 1+R}=-i\varepsilon^{-1/2} 
 {1-i\sigma(\varepsilon)\over 1+i\sigma(\varepsilon)}\; 
{\W '(-\varepsilon,0)\over \W (-\varepsilon,0)}\\ \nonumber 
=i\left({2\over\varepsilon}\right)^{1/2}  
{1-i\sigma(\varepsilon)\over 1+i\sigma(\varepsilon)}\; 
\left|{\Gamma(3/4-i\varepsilon/2)\over\Gamma(1/4-i\varepsilon/2)}\right|, 
\ee 
where the prime denotes the  derivative of  
$\W (-\varepsilon,\xi)$ with respect to $\xi$, 
and we expressed $\W$ and $\W '$ at $\xi=0$ in terms of  
the $\Gamma$-function, 
see  Ref. \onlinecite{Ab64}. The reflection coefficient entering the  
boundary conditions for the Boltzmann equation is $r=|R|^2$. 
 
Close to the mode opening threshold one can put $\varepsilon=0$ in $\sigma$ 
and $\Gamma$. Far from the threshold one can use the asymptotic 
for $\Gamma(x+iy)$ at $y\rightarrow\infty$, see Ref. \onlinecite{Ab64}. 
  This gives 
 
\beq  
r=\left\{ \begin{array}{ll} 
1-c\varepsilon^{1/2},  & 
(\varepsilon\ll 1),  \\ 
(1/4)  e^{-2\pi\varepsilon}, &(\varepsilon\gg 1) 
\end{array} \right. \, .  
\label{rapp} 
\eeq  
Here $c=2\Gamma(1/4)/\Gamma(3/4)\approx 5.92$.

Consider an electron with energy $E$ at  the midpoint  between the  
thresholds $E_{0}$ and $E_{1}$, i.e. $E-E_{0}=\Delta/2$. 
For  $a\gg d$ the waveguide opening is adiabatic. In this case 
 $\delta\ll \Delta$, and for the chosen  energy $E$ we have 
$\varepsilon\gg 1$ and  $r$ is exponentially small. 
However for a non-adiabatic  opening, when $a\simeq d$, 
for the same energy $E$ we find  $\delta\simeq \Delta$. 
As a result  we have $\varepsilon\simeq 1$ and  $r\simeq 1$. 
 
Using these results we can estimate the derivatives $\alpha$ and $\beta$ 
which determine the even and odd currents Eqs. (\ref{rect})  
and (\ref{drag}). For channel openings which  are not specially 
designed, $a\simeq d$, and  
 a typical Fermi energy in the midpoint  between the  
thresholds $E_{0}$ and $E_{1}$ (at the center of quantized conductance 
plateau) will yield  $\alpha,\beta\simeq 1$.

\widetext  
\end{document}